\documentclass[english]{article}
\usepackage{pgf,pgfarrows,pgfnodes,pgfautomata,pgfheaps,pgfshade}
\usepackage{amsmath,amssymb}
\usepackage[utf8]{inputenc}
\usepackage{colortbl}
\usepackage{cancel}
\usepackage[T1]{fontenc}
\usepackage{xmpmulti}
\usepackage{animate}
\usepackage{verbatim}
\usepackage{graphicx}
\usepackage{babel}

\begin{document}
\begin{titlepage}
\thispagestyle{empty}

\bigskip

\begin{center}
\noindent{\Large \textbf
{The Twin Paradox Put to Rest}}\\

\vspace{0,5cm}

\noindent{G. Alencar ${}^{a}$\footnote{e-mail: geova@fisica.ufc.br}}

\vspace{0,5cm}
 {\it ${}^a$Departamento de F\'{\i}sica, Universidade Federal do Cear\'{a}-
Caixa Postal 6030, Campus do Pici, 60455-760, Fortaleza, Cear\'{a}, Brazil. 
 }

\end{center}

\vspace{0.3cm}

\begin{abstract}
In this letter, we give a simple and analytical solution to the twin paradox. It relies just on Lorentz transformations and does not involve accelerated frames or any kind of signals exchanged between the twins. We expect that with this we can put a dead end to this century-old problem. 

\end{abstract}
\end{titlepage}

\section*{The Twin Paradox Put to Rest}
The twin paradox, or clock paradox, is a century-old problem in Special Relativity. It has been continuously discussed since the standard formulation of Langevin in 1911 \cite{Langevin}. Einstein by himself found a qualitative solution based on general relativity but did not achieve consensus. In the '50s, a long debate about the theme was done\cite{DINGLE}. Though the correct answer has never been in question, many solutions have been found in the literature. Therefore the matter of how to explain the apparent paradox is far from settled. For a review see Ref \cite{Shuler}. 
 
Many solutions follow Einstein and appeal to the acceleration of the rocket\cite{Einstein}. Other solutions rely only on Special Relativity(SR) and our solution is in this class. In general, the SR solutions are of two kinds: one involves light signals sent from the traveler to the Earth-based twin and the other uses the relativity of simultaneity. In both cases, spacetime diagrams are used to solve the problem.  These are the solutions presented, for example, in the famous book of Taylor and Wheeler \cite{Taylor}. The solutions which appeal to acceleration use seem to suggest that General Relativity is necessary to solve the paradox. However, Special Relativity is consistent by itself and some solution must exist without this.  At some point, a way to avoid the acceleration of the twin was found\cite{Romer}. However, the last author did not give a simple and analytical solution. It came as a surprise to us when writing a book about relativity \cite{livro}, the lack of a simple, analytical and direct solution to the paradox. Beyond this, in general, the solutions are very evolved and unclear. Up to now,  there is no solution that, together,  does not involve any kind of acceleration, signals, or spacetime diagrams and is based only on the Lorentz transformations. Below we present our solution, which has all these properties. To avoid long explanations and to ensure simplicity, we will be very direct. 

Let us define the problem. We have two twins on Earth, Alice, and Bob. Bob will travel to planet ``Air" and come back to Earth.  The proper distance, from Earth to the planet Air, is measured by Alice and given by $L$. The relevant events are shown in Fig. \ref{two}. 
\begin{figure}[!h]
      \label{two}
        \centering
        \includegraphics[scale=0.9]{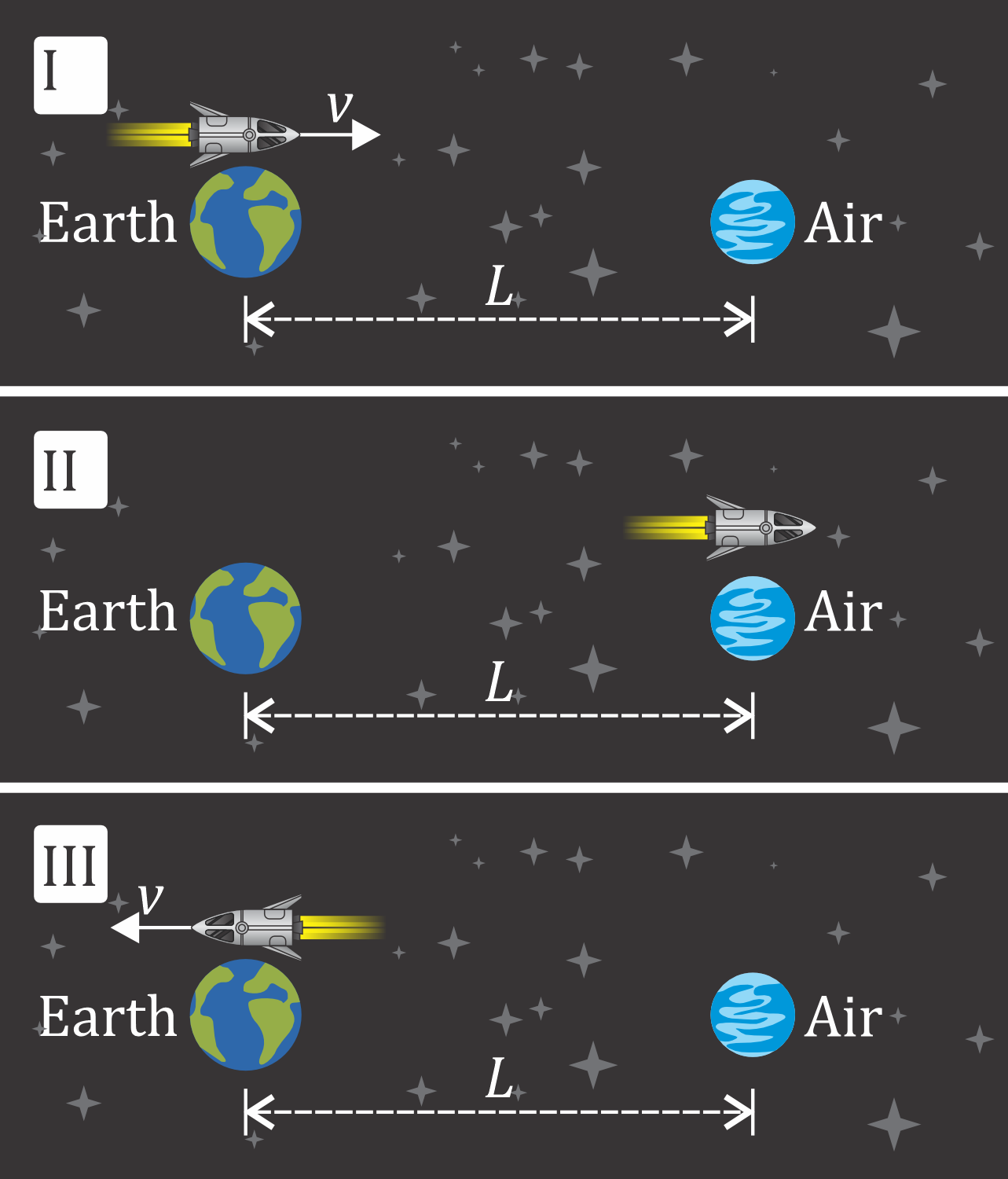}
        \caption{The relevant events as described by Alice: $I$) Bob departs from Earth, $II$) Bob arrives at ``Air" and $III$) Bob meets Alice at Earth}
        \label{two}
 \end{figure}
From Alice's clock viewpoint, the time travel of Bob is given by
\begin{equation}\label{relogioAlice}
 \Delta T_A=2L/v,   
\end{equation}
 where $v$ is the velocity of the rocket. However, Alice will see Bob younger due to time dilatation 
$$
\Delta T_B=\frac{\Delta T_A}{\gamma}=\frac{2L}{v\gamma}, \gamma =\frac{1}{\sqrt{1-\frac{v^2}{c^2}}}
$$
where $\gamma$ is the Lorentz factor. What about Bob? From his viewpoint, due to Lorentz contraction, the distance from the planet Fire is given by $L/\gamma$. Therefore the time travel is given by 
\begin{equation}\label{tempobobbob}
\Delta T'_B=\frac{2L}{v\gamma}.
\end{equation}
So, both agree with the time travel of Bob. The apparent paradox is that, for Bob, the clock of Alice goes slow by the factor
\begin{equation} \label{tempoaliceparabob}
 \Delta T'_A=\frac{\Delta T'_B}{\gamma}=\frac{2L}{v\gamma^2}.   
\end{equation}
Therefore, when they meet, Alice's clock will show (Eq. (\ref{relogioAlice})) or $2L/(v\gamma^2)$ (Eq. (\ref{tempoaliceparabob}))? This is the paradox. Alice says that Bob will come back younger due to its velocity. According to Bob, Alice is the one with velocity and therefore will be younger. The point is: when they meet and compare their clocks, who is correct?

\begin{figure}[!h]
        \centering
        \includegraphics[scale=0.9]{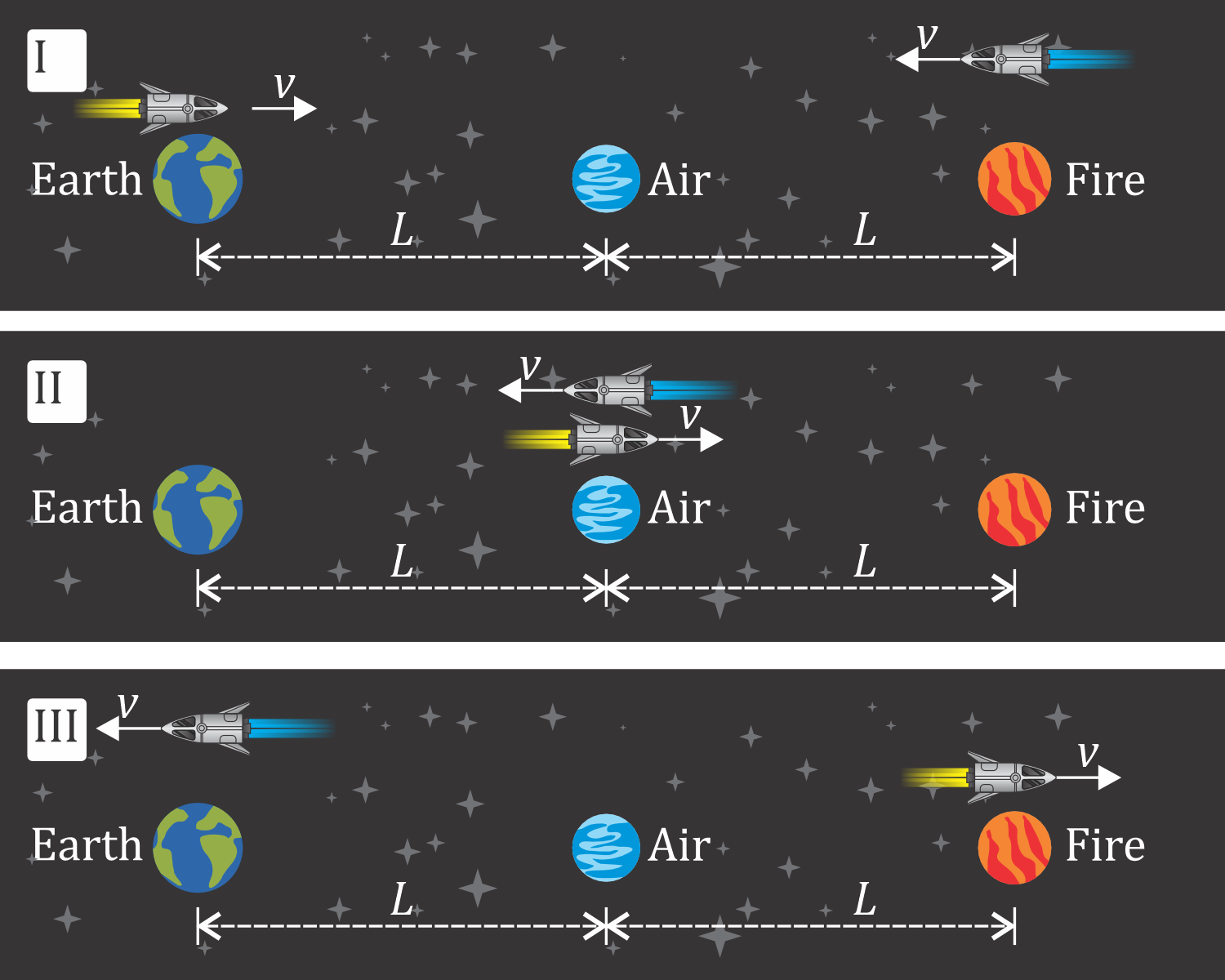}
        \caption{The relevant events as described by Alice: $I$) Bob and John depart from Earth and Fire, $II$) Bob and John arrive at ``Air" and $III$) John meets Alice on Earth and Bob arrives at Fire.}
           \label{three}
 \end{figure}

Bob is an astronaut, but Alice is a Physicist and says that Bob is not correct and he will be younger. The solution is the following. As said before,  we will avoid acceleration and use only Lorentz transformations. The way to do this was pointed out by Romer \cite{Romer}. We will use a slightly different configuration. To avoid accelerated frames, we will consider a third twin, John, traveling to Earth from a third planet, called ``Fire". The three planets are at rest with each other. Planet Air and Fire are at positions $L$ and  $2L$ from Earth. The relevant events are shown in Fig. \ref{three}.

To be more precise we will use the notation $I_E$ for the event $I$ at Earth in the reference frame of Alice. A prime, $I'_E$, will mean the same event in the reference frame of John. For Alice, the departures of Bob and John are simultaneous, with velocities $v$ and $-v$ respectively. Therefore, the events $I_E$ and $I_F$ are simultaneous for her. This way, John will arrive on earth at the same time as Bob arrives at Fire and events $III_E$ and $III_F$ are also simultaneous. Of course, Alice will say that John and Bob will be the same age. This is because the relative velocities, for her, are the same. This will simplify the problem a lot, since John has no acceleration and the clocks that will be compared are that of John and Alice, at event $III_E$. Of course, we could imagine another situation, in which Bob stops at Air and comes back to Earth. However, this is irrelevant since, for Alice, both of them must be younger and the same age. So, let us focus on Alice and John. 

As said above, for Alice the events $I_E$ and $I_F$ are simultaneous.  However,  for John, Earth and Fire have velocities $-v$. For him, and according to the Lorentz transformations
\begin{equation}\label{simultaneidade}
\Delta T'=\gamma(\Delta T+\frac{v}{c^2}\Delta x)=\gamma\frac{2Lv}{c^2}.
\end{equation}
In the second equality, we have used $\Delta T=0$ and $\Delta x=2L$. Therefore, the turning on of his clocks (event $I'_F$) and of Alice (event $I'_E$) are not simultaneous. The time difference between these events is given by
\begin{equation}\label{eq01}
I'_E-I'_F=\gamma\frac{2Lv}{c^2}.
\end{equation}
 This is the time difference in John's clock. Now, remember that, according to him, Alice's clock goes slow by the factor $1/\gamma$. Therefore, when John departs, the clock of Alice will be showing  $2Lv/c^2$. To discover what appears on her clock when John arrives on Earth, we must add the time of travel. As said above, from the viewpoint of Bob, this time goes slow and is given by Eq. \ref{tempoaliceparabob}. When we sum both times we arrive at
\begin{equation}\label{eq02}
\frac{2 Lv}{c^2}+\frac{2L}{v\gamma^2}=\frac{2L}{v}.
\end{equation}
Therefore, when John meets Alice, we will say that her clock will be showing $2L/v$. His clock, according to him, will show only the time travel, given by (\ref{tempobobbob}). We can state, so, that:
 \\
 \\
 \textbf{From the viewpoint of the two reference frames, both agree that the Alices's clock will be showing $2L/v$ and the one of John will be showing $2L/(v\gamma)$.}
 \\
 \\
 We conclude that, even though for John the clock of Alice goes slow, both agree that John, and therefore Bob, are younger. Due to the absence of simultaneity, for John, Alice's clock began to run before. For him, Alice is older. It is very interesting that, for John, the absence of simultaneity compensates exactly the time dilatation of Alice, in such a way that all of them agree that Alice will be older.

Finally, we should point out that Romer has achieved an important step by avoiding acceleration \cite{Romer}. However, in the last paper, the author just suggests that simultaneity \textbf{should} solve the problem, but does not present the exact calculation. As far as we know, the same attitude is present in all the solutions found in the literature.   The authors never show that simultaneity is \textbf{exactly} enough to solve the paradox. As far as we know, this is the first time that  Eqs. (\ref{eq01}) and (\ref{eq02}) are presented. Beyond being new, we believe that our solution deserves attention due to its simplicity and clearness since: a) It does not rely on accelerated frames, b) It does not depend on any kind of signals, c) It does not involve spacetime diagrams or any other complications and finally, the more important d) Eq. (\ref{eq02}) shows that simultaneity is exactly enough to solve the paradox. In fact, unlike the solutions found in the literature, it demands just two paragraphs! It is so simple that it is accessible to a high school student. Therefore we believe that it should be a standard solution contained in any relativity textbooks. We also expect that, with this, we can put a dead end to this century old problem. 

 \section*{Acknowledgments}

We acknowledge the financial support provided by the Conselho Nacional de Desenvolvimento Científico
e Tecnológico (CNPq) N 315568/2021-6 and Fundação Cearense de Apoio ao Desenvolvimento Científico  e
Tecnológico (FUNCAP) through PRONEM PNE0112- 00085.01.00/16.
\\
The authors have no conflicts to disclose.


\begin{thebibliography}{References}



%\cite{Langevin}
\bibitem{Langevin}
P.~Langevin,
%``The evolution of space and time,''
Scientia \textbf{10}, 31-54 (1911)
%9 citations counted in INSPIRE as of 31 Jan 2022



%\cite{DINGLE}
\bibitem{DINGLE}
DINGLE, H. 
``The ‘Clock Paradox’ of Relativity'',
Nature 179, 1242–1243 (1957).

%\cite{Shuler}
\bibitem{Shuler}
Shuler Jr., R.  
``The Twins Clock Paradox History and Perspectives", 
Journal of Modern Physics, Vol.5 No.12, 2014

%\cite{livro}
\bibitem{livro}
Alenca, G. ``Teoria da Relatividade para o Ensino Médio", To appear.

%\cite{Einstein}
\bibitem{Einstein}
Einstein, A. (2002) The Collected Papers of Albert Einstein, Volume 7: The Berlin Years: Writings, 1918-1921. (English Translation of Selected Texts). Princeton University Press, Princeton.

%\cite{Taylor}
\bibitem{Taylor}
Taylor, E. F., Wheeler, J. A., Wheeler, J. A. and Wheeler, J. A. 
 ``Spacetime physics: Introduction to special relativity, " 
New York: Freeman, 2002. Print.

%\cite{Romer}
\bibitem{Romer}
Robert H. Romer , "Twin Paradox in Special Relativity", American Journal of Physics 27, 131-135 (1959) https://doi.org/10.1119/1.1934783

\end{thebibliography}
\end{document}